# CONTEMPORARY APPROACH FOR SIMULATION AND COMPUTATION OF FLUID FLOWS IN CENTRIFUGAL HYDROMACHINES

*Alexey N. Kochevsky[1], Victor G. Nenya[2]*
[1] *Research Scientists, Department of Applied Fluid Mechanics*
[2] *Vice-Head of the Department of Basics of Machinery Designing*
*Sumy State University,*
*Rimsky-Korsakov str., 2, 40007, Sumy, Ukraine*
alkochevsky@mail.ru

***Abstract:*** *The article describes various aspects of mathematical modeling of fluid flows, both in general and with reference to hydraulic machinery. The article reviews historical development of corresponding methods of mathematical modeling. Implementation of these aspects in modern commercial CFD software tools is described together with advantages and disadvantages of implemented methods. The conclusion is drawn concerning possibilities of computation of fluid flows nowadays.*

## INTRODUCTION – HYSTORY OF DEVELOPMENT OF METHODS OF SIMULATION AND COMPUTATION

Just since the appearance of hydromachines (as well as all the other technical devices), their design engineers faced a problem of prediction of working parameters of a machine being designed before the drawings will be transferred to manufacture. When applying to centrifugal hydromachines the problem is still more complicated because their working parameters (head, efficiency, consumed power etc.) depend on the fluid flow pattern inside the hydraulic components. But this flow, due to the nature of the fluid medium, is so complex, that till now the only reliable method of research in fluid mechanics was the experiment. Only during the last years a significant progress was attained in creation of software tools for simulation of fluid flows. Nowadays, these tools (referred as CFD tools – Computational Fluid Dynamics) allow for simulation of fluid flows with so high reliability of the obtained results that the required volume of experimental research is often reduced to minimum.

Since the appearance of the first centrifugal pumps, the main reliable means for prediction of the performance curves of a pump are the formulas of similarity theory, provided the performance curves of a geometrically similar pump are known. As in real hydromachines the exact geometrical similarity seldom occurs, the empirical formulas were suggested allowing for taking into account these so called non-model changes [e.g., 1].

When a hydraulic machine of new design is to be created and appropriate performance characteristics are not available, till recently it was very difficult to predict its head and efficiency at the design stage. Prediction of performance characteristics for a new pump or turbine with proper precision is possible only by computing the fluid flow pattern inside its hydraulic components. General case of motion of fluid medium is described by Navier – Stokes equations. The solution of these equations, due to their extreme complexity, till recently was possible only with substantial simplifications.

Since the sixties of previous century, after appearance and distribution of first computers, the methods of simulation were applied where viscosity of fluid was neglected, i.e., the fluid assumed ideal and the flow assumed potential. In the USSR, the method of Raukhman B. S. [2] was widely recognized. This method allowed for computation of velocities and pressure at the contours of profiles laying at axisymmetrical flow surfaces, in 2D formulation.

Approximately at the same time, numerical methods for solution of Prandtl equations were developed that describe viscous fluid flow in a boundary layer, also in 2-dimensional formulation [3]. The first turbulence models were developed still earlier. Corresponding review is presented in the books [4, 5, 6].

Then, more complex methods were developed, where fluid flow was assumed to be potential everywhere except for boundary layers near solid walls where it was treated as viscous [e.g., 7]. Relating to hydromachines, such researches were conducted firstly in 2-dimensional formulation [e.g., 8] and then in 3-dimensional one.

In 70-s years, the first methods were developed allowing for numerical solution of full Navier – Stokes and Reynolds equations, both in 2-dimensional and 3-dimensional formulation [6, 9]. After a number of years, both methods of solution of these equations and turbulence models used gradually improved [6], a number of papers were published that demonstrated their successful application in practice. These equations were successfully applied also for simulation of fluid flow inside a rotating impeller [10].

Finally, during the last decade, the best of the developed methods of simulation and computation of liquid and gas flows in the regions of arbitrary geometrical configuration, including hydraulic components of hydromachines, were implemented as commercial CFD software tools, and these software tools were widely distributed at the market. Judging from the publications in the leading international journals on fluid dynamics, the most prominent successes were achieved by the groups of developers of CFX (Canada – England – Germany, www.software.aeat.com/cfx), STAR-CD (England, www.cd-adapco.com, www.adapco-online.com), Fluent (USA, www.fluent.com), Numeca (Belgium, www.numeca.be), FlowER (Ukraine, www.flower3d.org), etc. Profound documentation is supplied with these and other CFD software tools, making it possible for a qualified enough person to use these tools successfully for fluid flow computations, with no or minor technical assistance from the developers.

The present article describes the contemporary approach for simulation and computation of fluid flows in centrifugal hydromachines, as it is used in the CFD software tools mentioned above. This approach assumes the following sequence of actions: creation of geometrical model of the considered hydraulic components, generation of computational mesh, selection of the proper set of model flow equations, specification of boundary conditions, parameters of solution and other source data, running of solver and, finally, visualization and analysis of solution results. Below each of these items is described in more detail.

## CREATION OF GEOMETRICAL MODEL

The first stage of preparation of source data for computation of flow is creation of a solid geometrical model imitating the volume where the considered flow occurs. Because the hydraulic components of hydromachines (impeller, volute) are often of very complex shape, creation of their solid models is not a trivial problem.

At present, there are a number of software tools available at the market permitting to cope with this problem – SolidWorks (http://www.solidworks.com), ProEngineer (http://www.ptc.com), Unigraphics (http://www.eds.com/products/plm/unigraphics_nx), Catia (http://www.catia.ibm.com), T-Flex (Russia, http://www.tflex.com, http://www.tflex.ru), etc. After creation of the solid model with any of these software tools, they provide the possibility to save it in any of widely used file formats – VRML, IGES, STL. Thus, the complex surface of the solid model is approximated by a number of plane facets (cells), and these files contain nodal coordinates of these facets. Additionally, the CFD software tools listed in the introduction have their own means for creation of a geometrical model: CFX-Build in the package CFX, PROSTAR in the package STAR-CD, GAMBIT in the package Fluent.

The bladed components of hydraulic machines are of specific shape, and for creation of their geometrical models specialized software tools were developed, in particular, CFX-BladeGen. By default, the window of this software tool is split into 4 views. These views represent correspondingly the meridional projection, blade profile (at a given section), dependence of angle of incidence and spanning angle on axial coordinate and dependence of profile thickness on axial coordinate [11, 12]. Software interface of this tool allows also for representing of a lot of other important views, in particular, downstream variation of throat area and axonometry of the created solid model. The created model can be saved in any of the widely used file formats and transmitted further for generation of the computational mesh.

## GENERATION OF COMPUTATIONAL MESH

Generation of computational mesh is the process of splitting the computational domain into a set of discrete cells. The grid cells are polyhedrons, usually tetrahedrons, hexahedrons, prisms or pyramids (Fig. 1). Edges of these cells form lines of the computational mesh. The points located at the edges or in the center of a cell are grid nodes. As a result of numerical solution of model equations for fluid flow, the sought flow parameters are obtained just at the grid nodes.

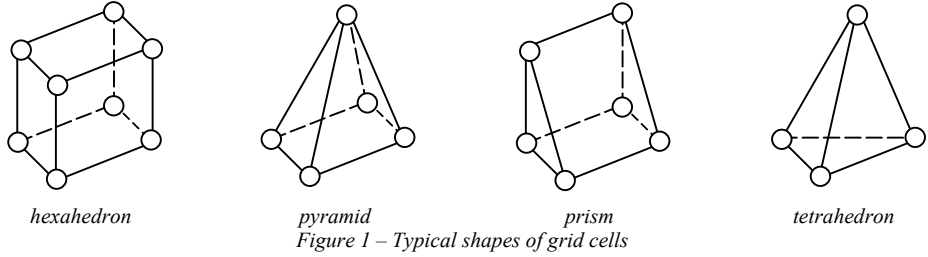

*hexahedron*    *pyramid*    *prism*    *tetrahedron*
Figure 1 – Typical shapes of grid cells

The main requirement for the computational mesh – it should be fine enough to resolve the physical effects occurring inside the computational domain. In order to achieve the uniform precision of solution, the grid nodes should be located denser at places where sharp changes of flow parameters occur, in particular, near the walls. Besides, when generating the mesh, obtaining of excessively stretched or skewed cells should be avoided. Presence of such cells can significantly embarrass obtaining of convergent solution.

One distinguishes structured and unstructured grids. In unstructured computational grids, the grid nodes are spreaded through the space by random way, according to the specified density law of node spreading. It makes possible to generate a mesh inside a domain of arbitrary geometrical complexity. But the discrete counterparts of model flow equations for such grids are cumbersome. In order to generate a structured grid, the computational domain is splitted into blocks according to some user-specified topology. The computational grids are generated inside each of the blocks, and the numbers of a 3-dimensional array can reference the grid nodes. Application of such a grid allows for composition of the most efficient solution algorithms [13].

As a rule, the leading CFD software packages for fluid flow solutions have their own grid generators (CFX-Build in CFX, pro-am in STAR-CD, GAMBIT and TGrid in Fluent, IGG Multi-blocks and IGG AutoGrid in Numeca). The software tool CFX-TurboGrid is designed for generation of high quality computational grids specially for bladed hydraulic components of hydromachines, allowing also for simulation of flow in the gap between rotor and stator parts [14]. The software tool G/Turbo has the same function in the package Fluent. Besides, the widely known grid generator ICEM CFD (http://www.icemcfd.com) is worth to be mentioned here. It is able to generate computational grids in the regions of arbitrary complexity and save them in formats compatible with each of CFD software packages mentioned above.

In order to prove that the obtained solution is grid independent, it is necessary to perform computations at several computational grids differing in the number of nodes. Thus, it is possible to determine the number of nodes, starting with which the solution almost will not change with further increase of this number.

## MODEL FLOW EQUATIONS

In the modern CFD software tools, the computation of liquid or gas flow is performed by numerical solution of system of equations that describe the most general case of movement of fluid medium. These equations are of Navier – Stokes (1) and continuity (2):

$$\frac{\partial}{\partial t}(\rho u_i) + \frac{\partial}{\partial x_j}(\rho u_i u_j) = -\frac{\partial p}{\partial x_i} + \frac{\partial}{\partial x_j}\left[\mu\left(\frac{\partial u_i}{\partial x_j} + \frac{\partial u_j}{\partial x_i}\right)\right] + f_i, \qquad (1)$$

$$\frac{\partial \rho}{\partial t} + \frac{\partial}{\partial x_j}(\rho u_j) = 0. \qquad (2)$$

For these equations, a brief form of record is used here. The summation on the same indices is assumed, $i, j = 1 \ldots 3$, $x_1, x_2, x_3$ – coordinate axes, $t$ – time. Full form of record for these equations in curvilinear coordinate system is presented, e.g., in [15]. The term $f_i$ expresses the action of body forces.

In this set of 4 equations, independent parameters being sought are 3 components of velocity $u_1$, $u_2$, $u_3$ and pressure $p$. Density $\rho$ of liquid as well as gas under velocities below 0.3 of Mach number is assumed to be constant.

Flows in rotating impellers of hydromachines are considered in the relative frame of reference. The term $f_i$ at the right-hand side of equations (1) expresses action of centrifugal and Coriolis forces:

$$\overline{f}_i = -\rho\left(2\overline{\omega}\times\overline{u} + \overline{\omega}\times(\overline{\omega}\times\overline{r})\right),$$

here, $\overline{\omega}$ is rotation speed, $\overline{r}$ is location vector (its length is equal to the distance from a given point to the axis of rotation).

The boundary conditions are posed usually as follows. Zero velocities are set at all the solid walls. At the inlet section, the distribution of all the velocity components is specified. At the outlet section, first derivatives of velocity components (in the direction of flow) are assumed to be zero. In practice, if velocity at the inlet section is approximately uniform, the user specifies only the average velocity (or flow rate). At the outlet section, the user usually does not specify anything, assuming the outlet section is located far enough from the regions of intensive flow transformation. The pressure is present in the equations (1) only in first derivatives, thus, the user needs to specify pressure only at any arbitrarily selected node of the computational domain.

As a rule, flows in centrifugal hydromachines are turbulent. Direct modeling of turbulent flows by numerical solving of Navier – Stokes equations, written for instant velocities, is still extremely difficult. Besides, as a rule, of interest are usually time averaged and not instant velocity values. Thus, for analysis of turbulent flows, instead of Navier – Stokes equations (1), Reynolds equations (3) are used:

$$\frac{\partial}{\partial t}(\rho\overline{u}_i) + \frac{\partial}{\partial x_j}(\rho\overline{u}_i\overline{u}_j) + \frac{\partial}{\partial x_j}(\rho\overline{u'_i u'_j}) = -\frac{\partial p}{\partial x_i} + \frac{\partial}{\partial x_j}\left[\mu\left(\frac{\partial \overline{u}_i}{\partial x_j} + \frac{\partial \overline{u}_j}{\partial x_i}\right)\right] + f_i, \quad (3)$$

where $\overline{u}_1$, $\overline{u}_2$, $\overline{u}_3$ – time averaged velocity components,

$\overline{u'_1}$, $\overline{u'_2}$, $\overline{u'_3}$ – fluctuating velocity components.

Different turbulence models are used for closure of these equations. These models are reviewed in the next chapter.

Besides those equations presented here, a number of other model equations are implemented in the leading CFD software tools. These model equations allow for simulation and computation of compressible flows (sub-, trans- and supersonic), flows with heat transfer (including transfer by radiation), flows with cavitation, flows of a mixture of several fluids, multiphase flows, flows with chemical reactions and combustion, etc. In general, tendency of development of the leading CFD software tools is implementation of a set of mathematical models in each of them, allowing for simulation of all the physical phenomena that may occur in practice as full as possible. A user turns on the necessary model equations while setting a problem, in several mouse clicks, and then specifies relevant boundary conditions and other required data.

It is necessary sometime to solve problems where elastic deformation of the domain occurs due to the pressure imposed by fluid flow. In such cases, simulation of fluid flow and simulation of wall deformation (e.g., impeller blade) are to be performed conjointly. Among the software tools able to cope with such problems, we should mention Ansys (www.ansys.com). Ansys is one of the most respected software packages used for strength problems as well as for conjugate simulation of processes of different physical nature. The package Ansys contains own software tool Flotran intended for computation of liquid and gas flows. Besides, the packages CFX and STAR-CD have data format compatible with Ansys, allowing for solution of indicated problems by joint application of these packages.

## TURBULENCE MODELING

In the modern software tools for simulation of liquid and gas flows, a lot of different turbulence models are used. In this review, we mention the most approved and widespread of them.

*1. Eddy Viscosity Models*

These models use the Boussinesq's assumption. According to this assumption, the terms with fluctuating velocities $\left(\rho\overline{u'_i u'_j}\right)$ in equations (3) are related to the averaged flow parameters by the following expression:

$$\rho\overline{u'_i u'_j} = -\mu_t\left(\frac{\partial \overline{u_i}}{\partial x_j} + \frac{\partial \overline{u_j}}{\partial x_i}\right) + \frac{2}{3}\rho\delta_{ij}k, \qquad (4)$$

where $\mu_t$ is turbulent viscosity, $k = 0.5\left(\overline{u'_j u'_j}\right)$ is turbulent kinetic energy, $\delta_{ij} = 1$ at $i = j$, $\delta_{ij} = 0$ at $i \neq j$.

These models are described, in particular, in [4, 5, 6], and are the most economical turbulence models used for computations of liquid and gas flows. Disadvantage of these models is impossibility (or restricted possibility) to take into account prehistory of flow, i.e., impossibility to model turbulent energy transfers from the upstream fluid layers. Consequently, these models are seldom used for computation of complex turbomachinery flows, though, e.g., Baldwin – Lomax model was successfully used for computation of compressible flows in [16].

In the modern CFD software tools, these models are used for prompt approximate analysis of fluid flows. In particular, a model of this group is implemented in the CFD tool CFX-BladeGenPlus belonging to the package CFX.

*2. Models Supposing Solution of 2 Additional Differential Equations*

Models of this group also use the Boussinesq's assumption (4).

Until now, $k - \varepsilon$ turbulence model developed in 70-s years [17], as well as its modifications, is still widely used in modern CFD software tools. When this model is used, the system of equations of fluid motion is added by 2 differential equations that describe transfer of turbulent kinetic energy $k$ and turbulent dissipation rate $\varepsilon$.

$$\frac{\partial}{\partial t}(\rho k) + \frac{\partial}{\partial x_j}\left(\rho \overline{u_j} k\right) = \frac{\partial}{\partial x_j}\left(\Gamma_k \frac{\partial k}{\partial x_j}\right) + P_k - \rho\varepsilon, \qquad (5)$$

$$\frac{\partial}{\partial t}(\rho\varepsilon) + \frac{\partial}{\partial x_j}\left(\rho \overline{u_j} \varepsilon\right) = \frac{\partial}{\partial x_j}\left(\Gamma_\varepsilon \frac{\partial \varepsilon}{\partial x_j}\right) + \frac{\varepsilon}{k}(C_{\varepsilon 1} P_k - \rho C_{\varepsilon 2}\varepsilon), \qquad (6)$$

where the term $P_k = -\rho\overline{u'_i u'_j}\frac{\partial \overline{u_i}}{\partial x_j}$ expresses generation of energy $k$,

$$\Gamma_k = \mu + \frac{\mu_t}{\sigma_k}, \qquad \Gamma_\varepsilon = \mu + \frac{\mu_t}{\sigma_\varepsilon}.$$

Parameters $\varepsilon$ and $\mu_t$ are defined as follows:

$$\varepsilon = \frac{\mu}{\rho}\overline{\left(\frac{\partial u'_i}{\partial x_j}\right)^2}, \qquad \mu_t = \rho C_\mu \frac{k^2}{\varepsilon}.$$

According to [17], the constants of $k - \varepsilon$ model are as follows: $C_\mu = 0.09$, $C_{\varepsilon 1} = 1.44$, $C_{\varepsilon 2} = 1.92$, $\sigma_k = 1.0$, $\sigma_\varepsilon = 1.3$.

Some aspects related to this model are described below.

- Various computations show very rapid variation of parameters $k$ and $\varepsilon$ near solid walls. In order to resolve these variations adequately, very dense computational grid should be used. The following approach is often used instead. A thin region is allotted near walls where numerical solution of equations (5) and (6) is not performed. Instead of this, the parameters are computed by algebraic formulas that describe typical near-wall layers [18]. In the modern CFD tools, in particular, CFX-TASCflow, both approaches are implemented.

- It was demonstrated recently [19] that the results of computations obtained using $k - \varepsilon$ model may depend strongly on the distance from the walls to the near-wall grid nodes. A contradiction to a principle of mathematical modeling was found out. That principle declares grid independence of the results of computations when the number of grid nodes in the domain is large enough. It was shown that grid nodes nearest to the walls should be located at the boundary of viscous sub-layer. In order to assure this condition, so-called scalable wall functions were implemented in CFX-TASCflow. Thus, the CFD tool selects proper grid nodes to switch to near-wall functions, preventing the user from erroneous computations.

Disadvantages of the $k - \varepsilon$ model is poor precision of simulation of flow separation from smooth surfaces and the described above difficulties with near-wall computations. In order to overcome these difficulties, the $k - \omega$ turbulence model developed by Wilcox can be used [20]. This model is also often used in modern CFD software tools. In this model, the second modeling parameter, instead of $\varepsilon$, is turbulent frequency $\omega$. Transfer of $k$ and $\omega$ is modeled by the following equations:

$$\frac{\partial}{\partial t}(\rho k) + \frac{\partial}{\partial x_j}\left(\rho \overline{u_j} k\right) = \frac{\partial}{\partial x_j}\left(\Gamma_k \frac{\partial k}{\partial x_j}\right) + P_k - \rho \beta^* k \omega, \qquad (7)$$

$$\frac{\partial}{\partial t}(\rho \omega) + \frac{\partial}{\partial x_j}\left(\rho \overline{u_j} \omega\right) = \frac{\partial}{\partial x_j}\left(\Gamma_\omega \frac{\partial \omega}{\partial x_j}\right) + \alpha \frac{\omega}{k} P_k - \rho \beta \omega^2, \qquad (8)$$

where $\Gamma_\omega = \mu + \dfrac{\mu_t}{\sigma_\omega}$, $\omega = \varepsilon / k \beta^*$, $\mu_t = \rho\, k\, /\, \omega$.

Constants of $k - \omega$ models, [20]: $\beta^* = 0.09$, $\alpha = 5/9$, $\beta = 3/40$, $\sigma_k = 2$, $\sigma_\omega = 2$.

In turn, disadvantages of $k - \omega$ model, in relation to $k - \varepsilon$ model, is excessive dependence of computational results on the values of $\omega$ specified at the inlet [21]. In order to combine advantages of these models, Menter [22] has suggested a hybrid turbulence model he named BSL (Baseline Model). In this model, a blending function $F_1$ is used for gradual switching from the $k - \varepsilon$ model that works well in the flow core to the $k - \omega$ model that works well near walls. Equations (7) and (8) are multiplied by $F_1$ and added to the corresponding equations (5) and (6) multiplied by $(1 - F_1)$. $F_1$ changes gradually from unity at the walls to zero outside the boundary layers.

In the same paper, Menter has suggested another model by offering a new formula for calculation of the blending function $F_1$ and setting a limiter to the formulation of turbulent viscosity $\mu_t$. This has resulted, in particular, in more exact simulation of flow separation from smooth surfaces. The new model was named SST (Shear Stress Transport) and implemented in CFX-TASCflow as the most adequate model among models based upon 2 additional differential equations [23]. Last years, different researchers have published a number of papers (e.g., [16, 24]) with results of simulations obtained just with SST model.

At the same time we should note that the Boussinesq's assumption used in all the above listed turbulence models is, in fact, the assumption of turbulence isotropy. In some cases, e.g., in strongly swirling flows, anisotropy of turbulent fluctuations is sharply expressed, and all these models fail in proper prediction of such flow patterns.

*3. Reynolds Stress Models*

These models, also implemented in the leading CFD software tools, reflect deeper understanding of the nature of turbulence and provide more opportunity to model relating physical effects. In particular, these models may be successfully used for computation of strongly swirling flows.

RSM turbulence model (Reynolds Stress Model) can be based upon $k - \varepsilon$, $k - \omega$ or SST model. In addition to 2 differential equations of these models, RSM model supposes 6 more differential equations to be included in the system of model equations. These additional equations model transfer of each of 6 Reynolds stresses: $\left(\rho \overline{u'_1 u'_1}\right)$, $\left(\rho \overline{u'_2 u'_2}\right)$, $\left(\rho \overline{u'_3 u'_3}\right)$, $\left(\rho \overline{u'_1 u'_2}\right)$, $\left(\rho \overline{u'_1 u'_3}\right)$ and $\left(\rho \overline{u'_2 u'_3}\right)$. These stresses are then substituted in the equations (3) without use of the Boussinesq's assumption (4). Equations of RSM model were deduced mostly owing to works of Rotta, in particular, [25, 26]. Disadvantages of this model are substantial increase of computing time per one iteration, and above all, substantial difficulties with achieving of convergent solution.

Due to these reasons, another model, ASM (Algebraic Stress Model), is often used in practice instead of RSM model. In this model, unlike RSM, transfer of each of 6 Reynolds stresses is modeled by not differential, but algebraic equations. This approach was suggested by Rodi [27]. This equations form a system with matrix of size 6 x 6 and are solved jointly, and the results are also substituted into the equations (3).

*4. LES and DNS Models*

These models (LES is Large Eddy Simulation, DNS is Direct Numerical Simulation) are maybe the most complex among all the turbulence models now available. While observing on turbulent processes, scientists have noticed that fluid particles in turbulent flows are involved in fluctuating motion that can be imagined as superposition of fluctuations of very different intensity and frequency. Moreover, the larger part of turbulent energy belongs to large scale fluctuating motion, i.e. fluctuations of large amplitude. These models present an attempt to simulate and compute large scale fluctuating motion directly (DNS model – by direct solution of Navier – Stokes equations for instant velocities). Small scale fluctuating motion (with amplitude of fluctuations below grid cell size) is simulated by simple turbulence models here.

These models were included only in the latest versions of CFX, STAR-CD and Fluent. A number of papers were already published demonstrating successful application of these models for computing of different flows [e.g., 28].

## DISCRETIZATION OF MODEL EQUATIONS

As it is known, the main approaches for discretization of model equations are FDM (finite difference method), FEM (finite element method) and FVM (finite volume method). All these methods may be considered as particular cases of a more general approach known as the method of weighted residuals. FDM is maybe the simplest to understand, but its application is rather difficult at unstructured grids. FEM works equally successfully both at structured and unstructured grids and thus is convenient to be applied for regions of arbitrary geometrical complexity. An important advantage of FVM is ensuring of conservation of integral parameters (flow rate, momentum) at each of finite grid cells, not only when the computational grid is dense enough.

As a rule, modern CFD software tools use just the FVM based upon the FEM approach. This permits to combine the indicated advantages of these methods.

A separate problematic aspect is discretization of the term $\frac{\partial}{\partial x_j}\left(\rho \overline{u_i u_j}\right)$ that expresses the convection process. When this term is discretized in the common way, precision of computation results is reduced as grid lines diverge from flow lines. The reason for this is so called numerical diffusion consisting in too rapid smoothing of velocity pattern between neighboring shears of flow, especially when computing transonic flows with shock waves. Recently, in the papers [29, 30, etc.] special schemes for discretization of this term were suggested that take into account the computed direction of flow, and now these schemes were implemented in the modern CFD software tools. Application of these discretization schemes is very important also for computation of hydromachinery flows featured with swirl and reverse motion, especially at off-design conditions.

## ALGORITHM OF NUMERICAL SOLUTION OF MODEL FLOW EQUATIONS

The algorithm SIMPLE developed by Patankar [9, 31] was one of the first algorithms for numerical solution of Navier – Stokes (and Reynolds) equations. This algorithm (with some modifications) is still used in a number of leading CFD software tools (in particular, STAR-CD). Solution process starts from a certain initial approximation imposed as source data. As a result of each global iteration, after elapsing of corresponding time step, new values of velocities and pressure are obtained. Stationary solution, if exists, is reached after completion of a large enough number of iterations which correspond to large enough period of time.

Though differential model equations (2), (3), (5) and (6) (or (7) and (8)) form a system, according to the method SIMPLE, they are solved at each iteration separately, in succession. Each equation is reduced to the system of linear algebraic equations (SLAE). As a result, a rather small size of SLAE matrix is obtained, moreover, this matrix is of approximately similar structure for (almost) each of those differential model equations. Thus, algorithm of solution as a whole is (relatively) simple, but convergence rate of a basic version of the algorithm SIMPLE for a number of problems is rather low.

Since then, some numerical techniques were developed making the algorithm to be more complex but allowing for acceleration of convergence rate. These techniques were implemented, in particular, in CFX-TASCflow.

*1. Multigrid Approach*

Some global iterations are performed at coarser computational grids allowing for more rapid approach to the converged solution. During this process, a user need not to generate new meshes, the CFD solver tool performs the whole process automatically. The solver tool itself finds the most proper directions to make the grid coarser in order to accelerate the convergence rate at a current step and composes discrete analogues of model equations relative to the nodes of this coarser grid. This algorithm is described in more detail in [32].

*2. Combined Solution of Continuity and Momentum Equations*

According to the algorithm suggested in [33] and implemented in CFX-TASCflow, numerical solution of equations (2) and (3) within one iteration is performed not in succession, but combined. This leads to multiple increase in size of SLAE matrix, its structure and algorithm of its solution becomes more complex. Thus, computational time per iteration increases. Nevertheless, this approach is prospective due to significant increase of convergence rate of the algorithm as a whole, – in order to obtain the convergent solution, it requires lesser number of iterations. Other model equations, (5) and (6) (or (7) and (8)), like in algorithm SIMPLE, are solved within iteration in succession.

## FEATURES OF COMPUTATION OF FLUID FLOWS IN CENTRIFUGAL HYDROMACHINES

Centrifugal hydromachines feature with rotation of rotor parts relative to stator parts leading to constant change of geometrical configuration of computational domain. Additionally, this domain is usually very complex in shape (large number of blade-to-blade channels, absence of symmetry).

Nowadays, several approaches were developed for simulation of such flows. These approaches differ in used assumptions, required computational resources and adequacy of obtained results.

*1. Single Frame of Reference*

*a) Steady Boundary Conditions*

This approach allows for flow computation in a single machine component. Stationary frame of reference is used for analysis of flow in guide vanes, rotating frame of reference – for flow in impeller. A user specifies velocity distributions (axial and circumferential velocity) at the entrance to the component considered.

This approach is convenient for draft analysis of flow in isolated machine components and is implemented, in particular, in the software tool CFX-BladeGenPlus belonging to the package CFX. This approach provides the most economical simulation. However, the results depend on boundary conditions the user may not know exactly. Besides, this approach does not allow for taking into account interaction between neighboring machine components.

*b) Transient (Periodically Changing) Boundary Conditions*

Time-dependent boundary conditions at the inlet present an attempt to simulate wakes from the blades of upstream machine component. These wakes are simulated by velocity distribution rotating relative to the used frame of reference.

This approach requires more computational efforts in comparison with the previous one. The disadvantages of the previous approach are specific also for this approach, though less emphasized.

*2. Two (or More) Frames of Reference*

In hydromachinery, as a rule, mutual influence of rotating and stationary bladed components is of significant importance. In order to simulate the flow properly, the computational domain should encompass the whole physical hydraulic domain, including stationary and rotating blade-to-blade channels. In order to save computational resources, only one blade-to-blade channel per machine component is often introduced into the computational domain assuming that in other blade-to-blade channels the flow pattern repeats itself periodically. On the other hand, this gives the opportunity to use computational grid dense enough for simulation of flow in separate channels.

Process of flow solution is performed jointly: within stationary machine components – in stationary frame of reference, within rotating impellers – in the frame of reference rotating together with impellers. In other words, just velocities in relative motion are used in model equations (2), (3), (5) and (6) (or (7) and (8)). The user prepares geometrical configuration of stationary and rotating components separately. Then, for performing the solution process, he attaches corresponding domains along the interface surfaces specified by him to make the united domain. The user specifies interface surfaces arbitrarily. It is advisable to locate them approximately in the middle between the neighboring bladed components (Fig. 2).

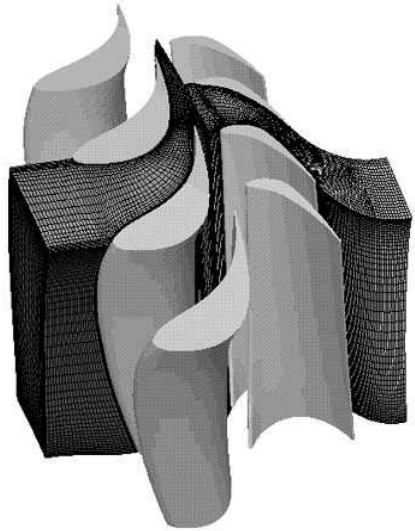

*Figure 2 – Typical computational domain, computational mesh and interface surface
in a centrifugal hydromachine ([http://www.adapco-online.com/feature_arts/rotflow1.html](http://www.adapco-online.com/feature_arts/rotflow1.html))*

Note, even if the number of blades in neighboring machine components is not equal or divisible (as a rule, it is just the case for hydromachines), the modern CFD software tools are smart enough to interpret correctly the attaching conditions at the interface surfaces.

In the modern CFD software tools, the following types of attaching conditions at the interface surfaces are implemented:

*а) Stage Averaging*

Flow parameters are circumferentially averaged at the sliding interface surface. In other words, this approach is based on the assumption that wakes from the blades of upstream component are smoothed away and the flow entering the downstream component is axisymmetric. This type of interface is appropriate when the distance between neighboring machinery components is large enough, for example, like in case of axial pumps.

*b) Frozen Rotor*

During the solution process, rotor components are kept fixed at a certain angular position relative to stator components. Flow parameters are not averaged at the sliding interface surface allowing for modeling of interaction of blade wakes from one component with blade-to-blade channel of next component. This type of interface is appropriate when the distance between neighboring machine components is small, for example, like in case of impeller and volute.

*в) True Transient*

With this approach, computation is performed without any simplifying assumptions. Relative angular position of rotor and stator bladed components is updated after each global iteration corresponding to one time step. This approach allows for most ultimate modeling of all transient effects occurring between the neighboring components. However, this approach is the most resource consuming and requires specification of very small time step. Solution process converges slowly. Besides, if number of blades in neighboring machine components is not divisible, a user should include into the solution domain all the blade-to-blade channels. Again, this increases the required resources and time of solution process many times.

In practice, as a rule, frozen rotor is used instead of true transient analysis. Performance parameters (head, efficiency) computed basing on the results of solution performed with these approaches, usually do not differ significantly.

## PARALLELIZATION OF COMPUTATIONS

The main purpose of simulation of flow in the hydraulic components of a pump or turbine is mostly prediction of its performance at the rated capacity. In order to perform such typical simulation, typical recommendations consist in building the computational domain consisting, e.g., of 1 vane-to-vane channel for guide vanes + 1 blade-to-blade channel of impeller; attaching condition at the interface surfaces is Stage Averaging or Frozen Rotor. According to practical experience, in order to reach the grid independent solution, the computational mesh altogether should include approximately 0.2 – 1 million of nodes. In this case, the solution process typically requires 0.2 – 1 Gbytes of computer memory and 0.2 – 1 day for a modern PC (e.g., [24]). Such a large number of nodes are typically just enough to satisfy the requirement of user guides of modern CFD software tools that at least 10 – 15 near-wall grid nodes should be located within the boundary layer in order to resolve it properly. When using coarser computational grids, the obtained results will be, most properly, also qualitatively correct, but not quantitatively precise.

The required memory and computation time increases multiply when it is necessary to simulate flow in several blade-to-blade channels or several stages, or when using complex turbulence models, or when capacity differ significantly from the rated one, or when physical processes of other nature are present, or when using the True Transient analysis, etc. It is possible to reduce multiply the duration of solution process by using several processors or several computers united in the net.

Parallel computations with multiprocessors are supported in each of the CFD software tools mentioned in the introduction. At the beginning of the solution process, a solver splits automatically the computational domain into separate subdomains. The size of each subdomain is proportional to performance of a corresponding processor. The general algorithm of computation works in such a way that each processor performs computation of flow only in "own" subdomain; after each global iteration the solutions obtained by all processor are "assembled" into the united solution. The algorithms of parallelization of computations used in modern CFD solvers provide acceleration of computations with the efficiency close to 100%. In other words, duration of the solution process is inversely proportional to the total performance of all the processors involved.

Nowadays, a large number of scientific papers relating to CFD is devoted to improvement of algorithms of parallelization of computations, e.g., see [34].

## VISUALIZATION AND ANALYSIS OF SOLUTION RESULTS

Process of computation of flow proceeds till the convergence criterion specified by the user will be satisfied, till the specified number of global iterations or until the user will interrupt it. By default, the modern CFD tools save the solution results as a file (or several files) containing all the information necessary for proceeding the solution process from the last state. It is possible to make a solver to save also intermediate states allowing for visualization of the process of transformation of flow pattern. Thus, the resulting file contains the coordinates of all nodes of the computational mesh and the values of main flow parameters in these nodes.

Interface of modern CFD software tools allows to represent the computational domain at the PC display and, e.g., paint this domain with different colors, in accordance with values of a computed variable – any standard variable or created by user. A user can create the formula of a new variable himself.

Typical visualization capabilities usually include: 2D graph, vector field, isolines and isosurfaces (equiscalar lines and surfaces), flood of different colors, animation of motion of fluid particles, etc. The computational domain together with the visualization pattern can be rotated, moved, scaled, etc., and this process can be written as an animated film.

The modern CFD tools also provide the capabilities of easy obtaining the performance parameters of flow, including those typical for hydromachinery: energy loss factor, head, consumed power, efficiency, torque, thrust, etc. A user can edit the formulas for computation of these parameters.

All these visualization capabilities allow viewing the computed flow pattern and analyzing the occurring physical effects (stagnation regions, reverse flows, etc.). It helps to understand in which way the geometrical parameters of the computational domain influence the performance parameters of flow.

Note that nowadays some specific visualization software tools are available at the market. These tools can read the files of results of CFD tools and provide still wider choice of visualization capabilities in comparison with that described above. These visualization tools, in particular, are TecPlot (http://www.amtec.com), ICEM CFD Visual 3 (http://www.icemcfd.com) etc.

## OPTIMIZATION

In the modern CFD software tools, the capabilities for optimization of geometrical parameters of the computational domain are also provided (in particular, the tool Optimus in the package FlowER). These tools provide interface where the user can specify the optimization problem, i.e., parameters of optimization, allowed range of their variation, other limitations and objective function, and specify the method of optimization. At each step of optimization, the CFD tool performs completely the flow computation in the domain with the corresponding set of values of geometrical parameters. Of course, the total duration of computation for this process is extremely large making this approach too complicated for common practice. Nevertheless, in future this approach is likely to become very powerful medium for designing of hydraulically perfect machine components.

Nowadays, an example of successful application of this approach is described, in particular, in [24]. In this paper, optimization of a turbine stage was conducted using the software packages FlowER (the tool Optimus) and CFX-TASCflow. Optimization was performed with limitations imposed as penalty functions. The computational mesh contained about 150 000 nodes per stage. In order to obtain the optimal solution using the Nelder – Mead method, 77 iterations were made which corresponded to 136 flow computations in the stage. Duration of solution of this problem was 12 days using PC Celeron with CPU of 1.3 MHz.

## CONCLUSION

The article presents a review of simulation and computation technology for fluid flows in centrifugal hydromachines using modern CFD software tools and describes the capabilities implemented in these tools. The review of modern publications demonstrates that the CFD tools mentioned above (CFX, STAR-CD, Fluent, Numeca, FlowER) allow for adequate modeling of complex physical phenomena occurring in fluid flows in hydraulic machine components and computing these flows within appropriate duration. These tools provide the user with convenient interface for input of source data and analysis of solution results. These tools also provide powerful capabilities for precise prediction of performance characteristics of hydraulic machines at the design stage allowing for saving resources for carrying out physical experiments.

**REFERENCES**

1. Biryukov A. I., Kochevsky N. N. Taking into Account of Influence of Non-model Changes upon the Performance Curves of a Centrifugal Stage // Bulletin of NTUU "KPI" – Kiev, 1999. – No. 36, Vol. 1. – P. 197-204. (in Ukrainian)
2. Raukhman B. S. Computation of Incompressible Fluid Flow around a Cascade of Profiles Taken at an Axisymmetric Surface in a Layer of Variable Thickness // Proceedings of Academy of Sciences of USSR, Fluid Mechanics. – 1971. – No. 1. – P. 83-89. (in Russian)
3. Rouleau W. T., Osterle J. F. The Application of Finite Difference Methods to Boundary-Layer Type Flows // J. Aeronaut. Sci. – 1955. – Vol. 22. – P. 249-254.
4. Fedyayevsky K. K., Ginevsky A. S., Kolesnikov A. V. Computation of a Turbulent Boundary Layer of Incompressible Fluid. – Leningrad: Sudostroyenie, 1973. – 256 p. (in Russian)
5. Cebeci T., Smith A. M. O. Analysis of Turbulent Boundary Layers. – New York: Academic, 1974.
6. Anderson D. A., Tannehill J. C., Pletcher R. H. Computational Fluid Mechanics and Heat Transfer. – New York: Hemisphere Publishing Corporation, 1984.
7. Kwon O. K., Pletcher R. H. Prediction of Incompressible Separated Boundary Layers Including Viscous-Inviscid Interaction // ASME Journal of Fluids Engineering. – 1979. – Vol. 101. – P. 466-472.

8. Kostornoy S. D. Mathematical Modeling of Fluid Flow in the Bladed Hydromachines with the Purpose of Determination of Their Hydrodynamic Characteristics for Analysis and Designing / Abstract of D.Sc. thesis – Kharkov: KhPI, 1992. – 35 p. (in Russian)
9. Patankar S. V., Spalding D. B. A Calculation Procedure for Heat, Mass and Momentum Transfer in Three-dimensional Parabolic Flows // Int. J. Heat Mass Transfer. – 1972. – Vol. 15. – P. 1787-1806.
10. Hah C., Bryans A. C., Moussa Z., Tomsho M. E. Application of Viscous Flow Computations for the Aerodynamic Performance of a Backswept Impeller at Various Operating Conditions // Journal of Turbomachinery – July 1988. – Vol. 110. – P. 303-311.
11. Bache G. CFX-BladeGen Version 4.0 Reaches New Heights in Blade Design // CFX Update – Spring 2001. – No. 20. – P. 9.
12. Plutecki J., Skrzypacz J. Zastosowanie Specjalistycznego Programu CFX-BladeGen w Procesie Projektowania Pomp // Pompy Popmownie. – 2003. – № 1. – C. 35-37. (in Polish)
13. Rusanov A. V., Yershov S. V. A Method of Computation of 3D Turbulent Flows in the Flow Passages of Arbitrary Shape // Perfection of Turbo-Installations by Methods of Mathematical and Physical Simulation: Collection of Papers. – Kharkov: Institute of Mechanical Engineering Problems of National Academy of Sciences of Ukraine, 2003. – Vol. 1. – P. 132-136. (in Russian)
14. Doormaal J. F. New Grid Generation For Rotating Machinery // CFX Update – Autumn 2002. – No. 22. – P. 6.
15. Kochevsky A. N. Optimization of Geometrical Parameters of the Discharge Diffuser of High-Speed Pumps of "IMP-GV" Type / Dissertation for Ph.D. degree in Mech. Eng. – Sumy: Sumy State University, 2001. – 195 p. (in Ukrainian)
16. Lampart P., Swirydczuk J., Gardzilewicz A., Yershov S., Rusanov A. The Comparison of Performance of the Menter Shear Stress Transport and Baldwin-Lomax Models with Respect to CFD Prediction of Losses in HP Axial Turbine Stages // Technologies for Fluid/Thermal/Structural/Chemical Systems with Industrial Applications, ASME. – 2001. – Vol. 424-2. – P. 1-12.
17. Launder B. E., Spalding D. B. The Numerical Computation of Turbulent Flows // Comp. Meth. Appl. Mech. Eng. – 1974. – Vol. 3. – P. 269-289.
18. Patel V. C., Rodi W., Scheuerer G. Turbulence Models for Near-Wall and Low Reynolds Number Flows: A Review // AIAA Journal. – September, 1985. – Vol. 23, No. 9. – P. 1308-1319.
19. Grotjans H., Menter F. R. Wall Functions for General Application CFD Codes // In ECCOMAS 98 Proceedings of the Fourth European Computational Fluid Dynamics Conference: John Wiley & Sons, 1998. – P. 1112-1117.
20. Wilcox D. C. Multiscale Model for Turbulent Flows // In AIAA 24th Aerospace Meeting / American Institute of Aeronautics and Astronautics, 1986.
21. Menter F. R. Multiscale Model for Turbulent Flows // In 24th Fluid Dynamic Conference / American Institute of Aeronautics and Astronautics, 1993.
22. Menter F. R. Two-Equation Eddy-Viscosity Turbulence Models for Engineering Applications // AIAA Journal. – 1994. – Vol. 32, No. 8.
23. Menter F. R., Esch T. Advanced Turbulence Modelling in CFX // CFX Update – Spring 2001. – No. 20. – P. 4-5.
24. Chupin P. V., Karelin D. V., Starkov R. Y., Shmotin Y. N., Yershov S. V., Rusanov A. V. Optimization of a Turbine Stage of HTD Using the Software Packages FlowER-Optimus and CFX-TASCflow // Perfection of Turbo-Installations by Methods of Mathematical and Physical Simulation: Collection of Papers. – Kharkov: Institute of Mechanical Engineering Problems of National Academy of Sciences of Ukraine, 2003. – Vol. 1. – P. 193-197. (in Russian)
25. Rotta J. C. Statistische Teorie nichthomogener Turbulenz // Zeitschrift für Physik. – 1951. – Vol. 129. – P. 547-572.
26. Rotta J. C. Turbulente Strömungen – B. G. Teubner Stuttgart, 1972.
27. Rodi W. A New Algebraic Relation for Calculation of the Reynolds Stresses // Zeitschrift für Angewandte Mathematik und Mechanik. – 1976. – Vol. 56. – P. 219-312.
28. Slitenko A. F., Gurinov A. A. Numerical Research of Influence of the Relative Pitch upon the Air-Dynamic Characteristics of the Turbine Cascade with Profiles N4U1 // Perfection of Turbo-Installations by Methods of Mathematical and Physical Simulation: Collection of Papers. – Kharkov: Institute of Mechanical Engineering Problems of National Academy of Sciences of Ukraine, 2003. – Vol. 1. – P. 149-154. (in Russian)
29. Raithby G. D., Torrance K. E. Upstream-weighted Differencing Schemes and Their Application to Elliptic Problems Involving Fluid Flow // Computational Fluids. – 1974. – Vol. 8, No. 12. – P. 191-206.
30. Raithby G. D. Skew Upstream Differencing Schemes for Problems Involving Fluid Flow // Computational Methods for Applied Mechanical Engineering. – 1976. – Vol. 9. – P. 153-164.
31. Patankar S. Numerical Heat Transfer and Fluid Flow. – New York: Hemisphere Publishing Corporation, 1980.
32. Hutchinson B. R., Raithby G. D. A Multigrid Method Based on the Additive Correction Strategy // Numerical Heat Transfer. – 1986. – Vol. 9. – P. 511-537.
33. Schneider G. E., Raw M. J. Control-Volume Finite Element Method for Heat Transfer and Fluid Flow Using Co-located Variables – 1. Computational Procedure // Numerical Heat Transfer. – 1987. – Vol. 11. – P. 363-390.
34. Yershov S. V., Rusanov A. V., Yershov D. S. A Universal Parallelization of Computation of 3D Viscous Flows for Computation Systems with Common and Distributed Memory // Perfection of Turbo-Installations by Methods of Mathematical and Physical Simulation: Collection of Papers. – Kharkov: Institute of Mechanical Engineering Problems of National Academy of Sciences of Ukraine, 2003. – Vol. 1. – P. 126-131. (in Russian)